\documentclass[aps,pre,twocolumn,showpacs,floatfix,superscriptaddress]{revtex4}
\usepackage{graphicx}
\usepackage{amsmath}
\usepackage{epsfig}
\newcommand{\on}{{\sc on}}
\newcommand{\off}{{\sc off}}
\newcommand{\bs}{\boldsymbol}

\begin{document}

\title{Attractors in continuous and Boolean networks} 

\author{Johannes Norrell}
\author{Bj{\"o}rn Samuelsson}
\author{Joshua E.\ S.\ Socolar}
\affiliation{
Physics Department and Center for Nonlinear and Complex Systems,
Duke University, Durham, North Carolina 27708}

\date{\today}

\begin{abstract}
  We study the stable attractors of a class of continuous dynamical
  systems that may be idealized as networks of Boolean elements, with
  the goal of determining which Boolean attractors, if any, are good
  approximations of the attractors of generic continuous systems.  We
  investigate the dynamics in simple rings and rings with one
  additional self-input.  An analysis of switching characteristics and
  pulse propagation explains the relation between attractors of the
  continuous systems and their Boolean approximations.  For simple rings,
  ``reliable'' Boolean attractors correspond to stable continuous
  attractors.  For networks with more complex logic, the qualitative
  features of continuous attractors are influenced by inherently
  non-Boolean characteristics of switching events.
\end{abstract}

\maketitle

\section{Introduction}
Complex networks of interacting elements arise in many biological,
chemical, sociological, and physical contexts.  An important example
is the network of interactions among proteins and DNA in a cell.  The
binding of certain proteins to each other and to promoter regions of
DNA can create a combinatorially complex logic of gene expression.  It
is tempting to think of transcriptional networks and similar examples
as effectively Boolean in nature.  In such a picture, genes are turned
``on'' and ``off'' in the presence of proteins produced when other
genes are turned on or off \cite{Davidson:06}, and many studies of the
fundamental principles governing networks of interacting genes have
focused Boolean models.  Recent work has highlighted distinctions in
the attractor structures as network architecture parameters are
varied, different distributions of Boolean functions are incorporated,
and/or different updating schemes are employed \cite{Aldana:03a,
  Aldana:03b, Greil:05, Moreira:05, Kesseli:06, Paul:06,
  Samuelsson:06}.

The Boolean models are generally understood to be idealized
representations of underlying continuous and perhaps stochastic
processes, so it is important to understand any artifacts introduced
in Boolean approximations.  Here we investigate some continuous,
deterministic systems motivated by models of transcriptional
interactions and designed to be good candidates for a Boolean
analysis.  The goal is to elucidate the most important effects that
may lead the continuous dynamics to differ qualitatively from
expectations based on the Boolean models.

We investigate the temporal structure and stability of the attractors
of continuous dynamical systems and those of the corresponding Boolean
models.  Two effects are found to be crucial for understanding the
structure of the continuous attractors: first, when an \on-\off\
symmetry of typical Boolean models is broken, the possibility of
stable pulse propagation down a chain can be lost; and second, the
memory of past inputs at a given node causes shifts in the temporal
spacing between multiple pulses on a feedback loop.  These effects
make it difficult to put the attractors of the continuous network into
correspondence with the attractors of their Boolean counterparts.

\begin{figure}[b]
  \includegraphics*[width=0.7\columnwidth]{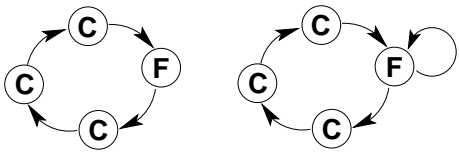}
\caption{\label{fig:rings} 
  A simple ring and a self-input ring.  {\bf\textsf C} indicates that
  the element copies its input; {\bf\textsf F} indicates that the
  element is either a copier or an inverter on the simple ring, or
  that it performs one of the two-input logic functions on the
  self-input element.}
\end{figure}

We study rings of $N$ elements (Fig.~\ref{fig:rings}) governed by
delay differential equations.  The delays are introduced to represent
intermediate steps in the process mediating the interactions between
elements.  We employ a form developed originally as a mean-field
description of the dynamics of transcription factor expression
\cite{Andrecut:06}, though for present purposes it simply provides
a generic model of elements with sigmoidal responses to their inputs:
\begin{align}
\label{eqn:continuous}
  {\dot x}_j(t) & = f_j\bs(x_{j-1}(t-\tau)\bs) - x_j(t)\,, \\
\label{eqn:f1in}
  f_j(x_i) & = 
    \eta_j\biggl(\frac{1+d_i^jx_i^{\nu}}{1+b_i^j x_i^{\nu}}\biggr)\,,
\end{align}
where $\eta_j$, $b_i^j$, and $d_i^j$ are constants and $\tau$ is a
time required for the signal produced by $x_{j-1}$ to reach $x_j$.
All subscripts are taken modulo $N$.  The parameters are chosen such
that the output of $f_j$ switches sharply between low and high values
as the input is smoothly varied.  We assume that all $x_j$ have the
same decay rate (chosen to be unity) and all arguments of $f_j$ have
the same time delay.  We also consider the effects of adding a
self-input to $x_1$, also with delay $\tau$:
\begin{align}
\label{eqn:selfinput}
  {\dot x}_1(t) & = f_1\bs(x_{0}(t-\tau),x_{1}(t-\tau)\bs) - x_1(t)\,, \\
\label{eqn:f2in}
  f_1(x_0,x_1) & = \eta_1\biggl(
      \frac{1+d_0^1 x_0^{\nu}+d_1^1 x_1^{\nu}+ d_{01}^1(x_0 x_1)^{\nu/2}}{\displaystyle
       1+b_0^1 x_0^{\nu}+b_1^1x_1^{\nu}+ b_{01}^1(x_0 x_1)^{\nu/2}}
\biggr)\,.
\end{align}
A Boolean idealization is obtained in the limit in which $f_j$ is a
step function and the decay term $-x_j(t)$ has an infinite
coefficient.

We find that continuous systems can exhibit stable oscillations in
cases where Boolean reasoning would suggest otherwise and that in some
cases where Boolean reasoning predicts a stable attractor, the
corresponding continuous attractor does not have the expected
structure.

\section{Boolean systems}
We begin by summarizing the known behavior of Boolean systems where
each $x_j$ is taken to be a Boolean variable and its dependence on its
inputs is specified by a Boolean function $F_j$.  A common choice is
to update all the elements synchronously, setting each $x_j$ at each
discrete time step to the value $F_j$ returned just after the previous
step.  Synchronously updated systems are easy to simulate, but are not
generic.  To avoid artifacts of synchronicity, Klemm and Bornholdt
\cite{Klemm:05a} present a class of autonomous Boolean systems running
in continuous time.  Here the output of each $F_j$ is fed through a
filter that delays the signal by a fixed time (analogous to our
$\tau$, which they set equal to 1) and cuts out short pulses.  (See
Appendix~\ref{app:autonomous} for details.)  These autonomous
networks have state cycles that correspond to the attractors of a
synchronously updated Boolean network.  In contrast to synchronous
Boolean networks, however, autonomous networks permit the study of
infinitesimal fluctuations in the timing of switching events.  One can
externally impose a delay in one switching event and see whether the
sequence of time intervals between switching events is restored by the
dynamics.  If all possible small time delays evolve back to the same
sequence of switching times, the state cycle of the autonomous system
is stable.  If, on the other hand, a subset of the switching times in
the cycle remain delayed compared to the others for some perturbation,
the state cycle is marginally stable.  There are no unstable state
cycles, because there is no way for an infinitesimal perturbation to
get amplified.  As a consequence, autonomous networks have an infinite
set of marginally stable cycles.

Klemm and Bornholdt coined the terms ``reliable'' and ``unreliable''
to denote attractors in the {\em synchronous} system that correspond
to stable and marginal cycles, respectively, in the corresponding
autonomous system \cite{Klemm:05a}. In accordance with their
convention, we use the term ``attractor'' for any periodic state-cycle
in a synchronously updated network.  Unreliable attractors are not
expected to be observed in real systems because errors in timing can
accumulate and eventually cause a transition to a different attractor.
One of our goals is to determine whether the set of reliable
attractors is in one-to-one correspondence with the set of attractors
of a continuous system.

The dynamics of simple Boolean rings have been well characterized
\cite{Flyvbjerg:88, Klemm:05a, Greil:05}.  In a simple ring, each
element either copies or inverts the state of its input.  A ring with
an even (odd) number of inverters is dynamically equivalent to a ring
with zero (exactly one) inverters since a pair of inverters can be
transformed to copiers by redefining the meaning of {\sc on} and {\sc
  off} for all elements between them.  For a ring with no inverters,
there are two fixed points; the states with all elements \on\ or all
elements \off.  For a ring with one inverter there is no fixed point
because at all times there must be at least one element whose value is
not consistent with its input.  We refer to this local inconsistency
as a ``kink'' and we identify the kink as ``positive'' (``negative'')
when the element's input is \on\ (\off).  A single kink traveling
around the ring forms a stable cycle with the kink changing its sign
each time it passes the inverter.  For synchronous updates, the
separation between kinks cannot change, so every state lies on an
attractor; there are no transients.  For the autonomous system, any
cycle with more than one kink is marginally stable because there is no
restoring mechanism for a perturbation in the time lag between two
kinks.  The multi-kink synchronous attractors are therefore
unreliable.

For rings in which element $1$ has a self-input in addition to its
input from element $0$, the attractor structure is nontrivial.
Without loss of generality, we assume that the remaining elements are
all copiers.  There are ten possible choices of the Boolean function
$F_1$ at element $1$ for which both inputs are relevant.  These
comprise five pairs that are related by an \on--\off\ symmetry.  For
two of these pairs, the only attractors are fixed points, so there are
only three nontrivial cases; (I) $F_1 =$ {\sc nor}, (II) $F_1 =$ {\sc
  xor}, and (III) $F_1 = (x_0$ {\sc and not} $x_1)$.

For the present discussion, we restrict attention to the case $N=4$.
In Case I, there is a single attractor that is (surprisingly)
unreliable.  In Cases II and III, the all-\off\ state is a fixed
point.  Case II also has a reliable cycle containing all 15 other
states.  Case III has two cyclic attractors, both unreliable: a
4-state cycle consisting of the state $1000$ and its cyclic
permutations; and a 2-state cycle consisting of the states $1010$ and
$0101$.

Our first important observation is that the definition of reliability
employed by Klemm and Bornholdt assumes the time delay for an element
is the same regardless of whether an input is turning \on\ or \off.
The breaking of this symmetry (see below) leads to different
propagation speeds for positive and negative kinks, which can destroy
or stabilize some marginally stable cycles.  For example, the
unreliable 4-cycle in Case III is stabilized if positive kinks move
faster than negative kinks.  Though the duration of the \on\ pulse
increases as it goes around the ring, it is cut back to $\tau$ each
time the pulse passes the self-input element.

\section{Analysis of continuous systems}
We now turn to the analysis of continuous systems described by
Eqs.~\eqref{eqn:continuous}--\eqref{eqn:f2in}.  We choose the Hill
coefficient (or cooperativity) $\nu=2$ corresponding to regulation
performed by dimers \cite{Andrecut:06}.  We fix $\eta$, $b$, and $d$
as follows in order to observe clearly identifiable \off\ and \on\
states.  For a single element that receives one input, we set
\begin{equation}
\label{eqn:singleinputparams}
  \begin{array}{llll}
  \eta=1, & b=0.001, & d=0.1 & {\textrm{\ for a copier}}; \\
  \eta=100, & b=0.1, & d=0 & {\textrm{\ for an inverter}}. 
  \end{array}
\end{equation}
With these choices, there are three fixed points for simple rings with
only {\sc copy} elements: the all-\off\ state $x=1.13$; the all-\on\ 
state $x=88.9$; and the unstable {\em switching value} $x=10.0$.  For
an {\sc invert} element, a steady \off\ input of $1.13$ produces
$x=88.7$ (close to the \on\ value); an \on\ input of $88.9$ produces
$x=0.13$ (close to the \off\ value); and $x=9.67$ is an unstable fixed
point.  For the self-input element, the values listed in
Table~\ref{tab:parameters} are chosen to represent the classes of
functions whose Boolean idealizations would be Cases I, II, and III
above.
\begin{table}[b] \label{tab:parameters}
\caption{Parameter values for self-input elements.}
\begin{tabular}{c|ccccccc}
\hline
 Case \rule[0pt]{0pt}{11pt} & $\eta$ & $b_0^1$ & $b_1^1$ & $b_{01}^1$ & $d_0^1$ & $d_1^1$ & $d_{01}^1$ \\
\hline
 I & $100$ & $0.1$ & $0.1$ & $0$ & $0$ & $0$ & $0$ \\
 II & 1 & $0.001$ & $0.001$ & $\;0.02\;$ & $\;0.1\;$ & $\;0.1\;$ & $\;0\;$ \\
 III & $1$ & $0.001$  & $0$ & $0.02$ & $0.1$ & $0$ & $0$ \\
\hline
\end{tabular}
\end{table}

To study the propagation of positive and negative kinks, we derive the
time it takes for a single kink to pass from one element to the next
in a chain of single-input elements.  The expression levels are
denoted by $x_0(t),x_1(t),x_2(t),\ldots$, where $x_0(t)$ serves as an
input and the other elements obey Eq.~\eqref{eqn:continuous}.  
For initial conditions, we
assume that $x_j(t)$ has the fixed value $\alpha_j$ for all $t\le0$,
with $\alpha_j=f_j(\alpha_{j-1})$ for $j=1,2,\ldots$.

If $x_0$ switches values, approaching a constant value $\beta_0$ at long times, each of
the $x_j$ will eventually approach a new value $\beta_j$.  For
mathematical convenience, we define rescaled quantities ${\hat x}_j(t)
\equiv [x_j(t) - \beta_j]/(\alpha_j - \beta_j)$ and ${\hat f}_j(x)
\equiv [f_j(x) - \beta_j]/(\alpha_j - \beta_j)$ with the properties
${\hat x}_j(0) = 1$ and ${\hat x}_j(t)\rightarrow 0$ for large $t$.
We further define a specific time $s_j$ associated with the switch
from $\alpha_j$ to $\beta_j$ by the formula
\begin{equation}
\label{eqn:switchingdef}
  s_j \equiv \int_0^\infty dt\,{\hat x}_j(t).
\end{equation}

The formal solution of Eq.~\eqref{eqn:continuous} is
\begin{equation}
\label{eqn:dynamicsintegral}
  x_j(t) = \int_{-\infty}^tdu\,f_j\bs(x_{j-1}(u-\tau)\bs)\,e^{u-t}
\end{equation}
for all $j\neq 0$.  Substituting into Eq.~\eqref{eqn:switchingdef} and
subtracting $s_{j-1}$ from both sides yields an expression for the
time delay between the switching of elements $j-1$ and $j$:
\begin{equation}
\label{eqn:switchingdiff}
  s_j - s_{j-1} = \tau + 1
    + \int_0^\infty dt\,[{\hat f}_j\bs(x_{j-1}(t)\bs)-{\hat x}_{j-1}(t)].
\end{equation}
This delay is the sum of the explicit delay $\tau$, an intrinsic delay
of unity associated with the unit coefficient of the $-x_j$ term of
Eq.~\eqref{eqn:continuous}, and an additional term that depends on the
details of the input function ${\hat f}_j$.

For chains of identical copiers with the parameters
specified above, it is straightforward to show that the additional
delay is positive (negative) for negative (positive) kinks and
therefore that positive kinks will propagate faster, which implies
that an \on\ pulse will expand in width and an \off\ pulse will
shrink and disappear.  By changing the parameters $\eta$, $b$, and
$d$, it is possible to reverse this situation, but arranging for
precisely equal propagation speeds requires fine tuning.

\begin{figure}[t]
  \includegraphics*[width=\columnwidth]{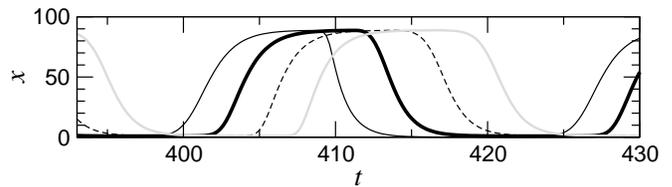}
\caption{\label{fig:kinkspeeds}
  Propagation of a single kink around a ring of four elements with one
  inverter and $\tau=2$.  The thick, thin, dashed, and grey lines
  represent element 1, 2, 3, and 0, respectively.  Note that
  propagation is faster when the kink is positive and slower after the
  inverter converts it to a negative kink.  }
\end{figure}

In the numerical simulation of Fig.~\ref{fig:kinkspeeds} it is clear
that the kink moves faster in its positive form than in its negative
form.  (To bring out the asymmetry, this figure was made for $\tau =
2$, a relatively short time.  Though the propagation time from one
site to the next depends on $\tau$, the difference between propagation
times for positive and negative kinks does not.)  The asymmetry is
also present in the continuous two-element rings in
Ref.~\cite{Klemm:05b} and in the repressilator simulation of
Ref.~\cite{Elowitz:00}.  The electronic model with step function
switching in Ref.~\cite{Glass:05}, however, does not break the
symmetry for the case studied, in which the switching level is halfway
between the \on\ and \off\ voltages.

Until now, we have neglected the fact that the propagation speed of a
kink through a given element is influenced by kinks that have
previously passed through.  
Consider now an \on\ pulse consisting of a positive kink followed by
a negative kink.
From Eq.~\eqref{eqn:dynamicsintegral}, we
see that $x_j(t)$ has an exponentially decaying memory of the events
in $x_{j-1}(t)$.  Because $f_j(x)$ is monotonic, the memory of low
values of $x_{j-1}(t)$ before the pulse in $x_{j-1}(t)$ will speed
up the effect of the trailing edge.  The interaction
shortens the pulse duration at $x_j(t)$, and monotonicity of $f_j$
ensures that the effect is stronger for shorter pulses.  Similar
reasoning shows that an \off\ pulse will also be shorter than the
time between kinks one would predict  
from Eq.~\eqref{eqn:switchingdiff} alone.

In a chain of copiers, different propagation speeds of
positive and negative kinks will lengthen (shorten) a traveling \on\
(\off) pulse.  In a chain that contains inverters arranged such that a
single pulse spends equal amounts of time in its \on\ and \off\
configurations, the asymmetry between positive and negative kinks
alone may not change the average pulse duration, but the pulse would
still be shortened by the memory effect.

Because the strength of the memory effect increases as two kinks
approach each other, a pulse or other sequence of kinks cannot
propagate stably on a chain of single-input elements --- only a single
kink can have a stable shape as it advances.  For simple rings, then,
stable attractors must have only zero or one kink.  These two
possibilities correspond precisely to the fixed points and single-kink cycles that
are the reliable attractors of the corresponding Boolean systems.

For rings with a self-input (Fig.~\ref{fig:rings}), the situation is
more complicated.  First, note that the analysis of pulse propagation
gives a quantitative measure of the asymmetry discussed above.  The
asymmetry enables some attractors classified as unreliable by the
(symmetric) Boolean analysis to be stable in the continuous system.
Second, memory effects associated with multiple inputs to a single
element can lead to repulsion between pulses and stabilization of new
attractors
related to synchronous Boolean ones but with shifts in the timing
between pulses.  

\section{Numerical results}
\begin{figure}[t]
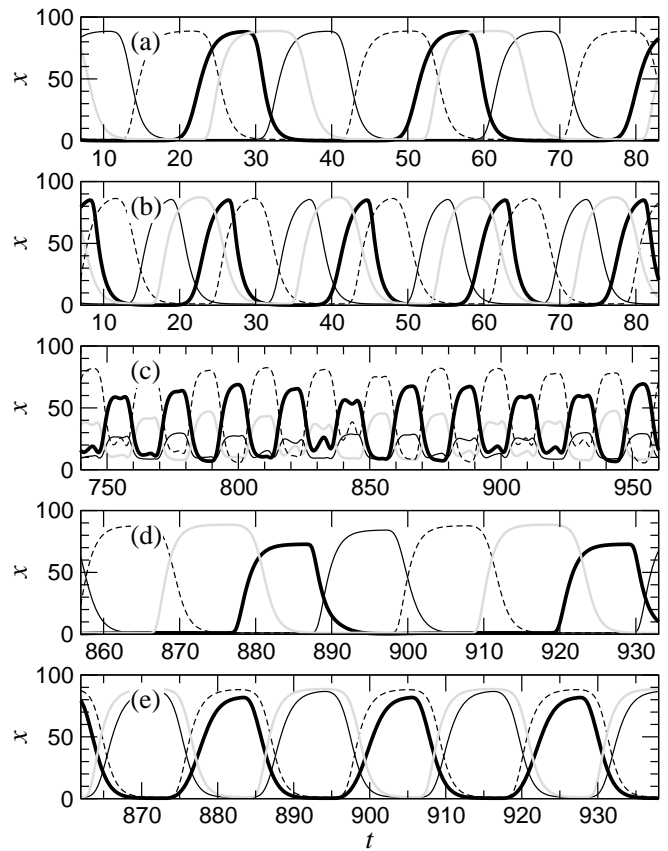

  \includegraphics*[width=\columnwidth]{nor.eps} \\
  \includegraphics*[width=\columnwidth]{nor3.eps} \\
  \includegraphics*[width=\columnwidth]{xor.eps} \\
  \includegraphics*[width=\columnwidth]{singlePulse.eps} \\
  \includegraphics*[width=\columnwidth]{doublePulse.eps}
\caption{\label{fig:timeseries}
  Time series for Cases I (a and b), II (c), and III (d and e), with
  $\tau=10$.  The thick, thin, dashed, and grey lines represent
  element 1, 2, 3, and 0, respectively.  Case I shows (a)
  non-synchronous timing stabilization due to memory effects (with
  time axis labels shifted by $10^6$ to account for a long transient)
  and (b) a stable attractor that has no analogue in the corresponding
  synchronous Boolean
  network.  Case II shows (c) an inherently non-Boolean attractor.  Case
  III shows (d) single pulse and (e) double pulse attractors, both of
  which are unreliable in a Boolean network.}
\end{figure}

We studied the three cases numerically using an integration scheme
that takes advantage of the structure of
Eq.~\eqref{eqn:dynamicsintegral} as described in
Appendix~\ref{app:methods}.  In Case I, most initial conditions
lead first to a long transient that corresponds to the unreliable
attractor observed in the synchronous Boolean network.  The transient
eventually gives way to a stable attractor of
Fig.~\ref{fig:timeseries}(a), which has two pulses separated in time
by approximately $(5/2)(\tau + 1)$, an example of the stabilization of
an intermediate inter-pulse interval where the synchronous Boolean
attractor has pulses separated in time by alternating intervals of $2$
and $3$ time steps.  We note in passing that the path by which the
continuous system arrives at the attractor is rather robust: for a
wide range of initial conditions, the system goes to the Boolean-like
transient in a short time and then gradually shifts to the attractor.
In some cases, however, we observe the attractor shown in
Fig.~\ref{fig:timeseries}(b), which has three pulses with a time
separation of $(5/3)(\tau + 1)$ between each pair.  This attractor
corresponds to a marginally stable cycle of the autonomous Boolean
network, but the timing is incompatible with the synchronous updating
scheme.  

We note that the appearance of attractor periods having
$5(\tau+1)$ as an integer multiple can be understood analytically.
Kaufman and Drossel have analyzed the possible attractor periods in
synchronous Boolean networks in rings with a single cross-link
\cite{Kaufman:05}.  A straightforward extension of their method to the
autonomous networks of our ring with a {\sc nor} self-input (Case I)
reveals that all cycles in the autonomous model must have a period of
the form $5/m$, where $m$ is an integer and the time unit is the time
required for a kink to advance through one element.  The corresponding
period in the continuous model is therefore $5(\tau +1)/m$.

In Case II, two attractors are observed: the fixed point and an
oscillatory attractor that appears to be aperiodic.  The latter
behavior is sensitive to intermediate variable values and is thus
inherently non-Boolean.  In Case III, we observe one fixed point and
two limit cycles, corresponding to each of the three Boolean
attractors, with no long transients.  The limit cycles are {\em
  unreliable} in the Boolean case but {\em stable} in the continuous
case --- examples of stabilization due to asymmetry in kink
propagation speeds and a pulse-shortening self-input element.  
This is the effect alluded to above at the end of the 
Boolean systems section.  Though the pulse has broadened in traveling
around the ring, a new trailing edge is generated by the self-input,
which occurs one delay time after the arrival of the leading edge.

The
difference in pulse height in the two limit cycles is due to the
dependence of $f_1(x_0, x_1)$ on small variations in $x_1$ when $x_0$
is nominally \on\ and $x_1$ is nominally \off.  The variations in the
\off-state of $x_1$ are caused by the delayed self-input that
suppresses the \off-state during a time of approximately $\tau$
following each pulse.  In the limit cycle with two pulses, the
suppression is still active when the next pulse starts.  This is not
the case for the single-pulse attractor and the slightly higher
\off-state leads to a suppression of the pulse height.

The above observations remain valid for all sufficiently large values
of $\tau$.  For small $\tau$, the systems studied have only fixed
point attractors.  As $\tau$ is increased, the oscillatory attractors
are born by means of saddle-node bifurcations of cycles
\cite{Strogatz:00} in Cases II and III.  In Case I, a subcritical
bifurcation to a state cycle with period near $3\times(5/2)(\tau+1)$
is followed by a subcritical, symmetry-restoring bifurcation with
period near $(5/2)(\tau+1)$.  The cycle with period near
$3\times(5/2)(\tau+1)$ has no direct correspondence to a cycle in the
autonomous Boolean network.

\section{Discussion}

Our study of simple and single-self-input rings has elucidated some
important non-Boolean features of continuous systems: (i) the
asymmetry in the reaction time of an element when an input switches
\on\ or \off; (ii) deviations from nominal \on\ and \off\ values; and
(iii) memory effects due to the exponential decay of variables to
their steady state values for fixed inputs.  These features are
crucial for the stabilization and destabilization of oscillations in
the systems we have investigated.  In simple rings, they lead to a set
of stable attractors that coincides precisely with the reliable
attractors identified by Klemm and Bornholdt \cite{Klemm:05a}.  When
more complex logic is introduced, as illustrated in here by adding a
self-input to one element in a ring, the stable continuous attractors
do not correspond to the reliable attractors.

These observations are important for generalizing the well-developed
theory of large random Boolean networks to generic systems.  We would
like to know, for example, whether large, complex networks exhibit a
well-defined dynamical phase transition similar to the ``order-chaos''
transition in ensembles of Boolean networks
\cite{Kauffman:69,Derrida:86a,Samuelsson:06}.  An important feature of
the Boolean network transition is the rapid scaling of the number of
attractors with system size in the disordered regime, which suggests
that we should try to understand the set of attractors of large and
complex continuous systems.

The non-Boolean effects discussed in this paper may also be directly
relevant for understanding cell cycle oscillations in yeast, where
there is evidence for a transcriptional oscillator and recent
proposals for the genes involved suggest a fundamental ring of four
elements with multiple feed-forward and feedback links
\cite{pramila:06}.  We have seen, for example, that a pulse of
activity may propagate stably in such networks even where Boolean
reasoning suggests otherwise.  In systems where distinct elements have
significantly different time delays, we expect additional deviations
from the dynamics in synchronous or autonomous Boolean networks.
Future studies along these lines should elucidate the behavior of
larger rings and more complex network structures.


\begin{acknowledgments}
We thank S.~Kauffman, A.~Ribeiro, and M.~Andrecut for stimulating
conversations and the U.~Calgary Inst.\ for
Biocomplexity and Informatics for hosting extended visits by
Samuelsson and Socolar.  This work was supported by the National
Science Foundation through Grant No.~PHY-0417372 and the Graduate
Research Fellowship Program.
\end{acknowledgments}

\appendix

\section{Autonomous dynamics}
\label{app:autonomous}

For completeness, we describe the implementation of autonomous Boolean
systems introduced by Klemm and Bornholdt \cite{Klemm:05a}.  The
important features are that switching times are determined by local
time delays and that repeated switching on times scales much shorter
than the delay time is suppressed.  There is no external clock and
no stochastic rule for determining update times.

As in Ref.~\cite{Klemm:05a}, we assume a time delay of $1$ unit
between the switching of a given node and the switching it induces in
nodes directly linked to it.  At each node, a low-pass filter is
assumed to suppress spikes of duration much shorter than $1$.  We let
$\epsilon$ be the minimum duration of an output pulse that passes
the filter (and require $0<\epsilon\ll1$).  If the inputs to a given
node switch at times that would lead to a pulse shorter than
$\epsilon$, the output from that node is assumed to stay constant
during that time.

For notational convenience, we write $F_j(t)$, suppressing its direct
dependence on its inputs.  We also let the Boolean values \off\ and
\on, respectively, correspond to the real values $0$ and $1$.  Then,
$x_j(t)$ is given by
\begin{align}\label{eqn:filter}
  x_j(t) &= \Theta\biggl[\frac{1}{2\epsilon}
            \int_{-\epsilon}^{\epsilon}d\delta\left(F_j(t-1+\delta) - \frac{1}{2}\right)
            \biggr]\,,
\end{align}
where $\Theta(x)$ is the Heaviside step function; $\Theta(x) = 0$ for
$x<0$ and $1$ for $x\ge0$.

When the timing of some switching events are perturbed
infinitesimally, $F_j$ may switch twice in rapid succession, which
would generate a positive or negative pulse of infinitesimal duration.
The filter of Eq.~\eqref{eqn:filter} suppresses such spikes, ensuring
that the number of switching events remains constant (for cycles in
which all spikes are longer than $\epsilon$ in duration) and
therefore that the stability of a state cycle is well defined with
respect to infinitesimal timing perturbations.

\section{Numerical integration method}
\label{app:methods}

Numerical integration of the time-delay equations
\eqref{eqn:continuous}--\eqref{eqn:f2in} was accomplished using a
fourth-order scheme based on the solution shown in
Eq.~\eqref{eqn:dynamicsintegral}.  To evolve the system through a time
step $h$, we define values of $x$ at the time points $t = nh$ for all
integer $n$ and define $m\equiv \tau/h$, where $\tau$ is the delay
time and $h$ is chosen such that $m$ is an integer.  To advance from
time step $n-1$ to $n$ we use the following formula (suppressing the
element index subscript $j$ for notational simplicity):
\begin{align}
\label{eqn:numerics}
  x_n & = 2h\left(\frac{1}{6}f_{n-m}+\frac{2}{3}e^{-h}f_{n-m-1}+\frac{1}{6}e^{-2h}f_{n-m-2}\right) \nonumber \\ 
     \  & \phantom{=}~ +\, e^{-2h}x_{n-2}\,,
\end{align}
where $f_n$ is the function defined by Eq.~\eqref{eqn:f1in}
[or~\eqref{eqn:f2in}] evaluated at time $t=nh$.  The integrator is
initialized with a desired value of $x_0$, where it is
assumed that $x_{n} = x_0$ for all $n<0$.  Integrations were carried
out using $h=0.1$.  Decreasing $h$ to $0.01$ had no noticeable effect
on the results.

Attractors were found by running from many (of order 50) different
initial conditions.  In some cases, such as the attractor shown in
Fig.~\ref{fig:timeseries}(b), it was necessary to arrange initial
conditions in which some nodes were artificially held at constant
values and released at different times.  The bifurcation structures as
a function of the time-delay parameter $\tau$ were determined by
performing integrations in which $h$ was increased or decreased very
slowly and observing transitions in the oscillation patterns of all
$x$'s.

\vfill

\bibliography{rings}

\begin{thebibliography}{19}
\expandafter\ifx\csname natexlab\endcsname\relax\def\natexlab#1{#1}\fi
\expandafter\ifx\csname bibnamefont\endcsname\relax
  \def\bibnamefont#1{#1}\fi
\expandafter\ifx\csname bibfnamefont\endcsname\relax
  \def\bibfnamefont#1{#1}\fi
\expandafter\ifx\csname citenamefont\endcsname\relax
  \def\citenamefont#1{#1}\fi
\expandafter\ifx\csname url\endcsname\relax
  \def\url#1{\texttt{#1}}\fi
\expandafter\ifx\csname urlprefix\endcsname\relax\def\urlprefix{URL }\fi
\providecommand{\bibinfo}[2]{#2}
\providecommand{\eprint}[2][]{\url{#2}}

\bibitem[{\citenamefont{Davidson}(2006)}]{Davidson:06}
\bibinfo{author}{\bibfnamefont{E.~H.} \bibnamefont{Davidson}},
  \emph{\bibinfo{title}{The regulatory genome: gene regulatory networks in
  development and evolution}} (\bibinfo{publisher}{Academic Press},
  \bibinfo{year}{2006}).

\bibitem[{\citenamefont{Aldana-Gonzalez
  et~al.}(2003)\citenamefont{Aldana-Gonzalez, Coppersmith, and
  Kadanoff}}]{Aldana:03a}
\bibinfo{author}{\bibfnamefont{M.}~\bibnamefont{Aldana-Gonzalez}},
  \bibinfo{author}{\bibfnamefont{S.}~\bibnamefont{Coppersmith}},
  \bibnamefont{and} \bibinfo{author}{\bibfnamefont{L.~P.}
  \bibnamefont{Kadanoff}}, \emph{\bibinfo{title}{Boolean Dynamics with Random
  Couplings {\normalfont in} {P}erspectives and Problems in Nonlinear
  Science{\normalfont, edited by E. Kaplan, J. E. Marsden and K. R.
  Sreenivasan}}} (\bibinfo{publisher}{Springer, New York},
  \bibinfo{year}{2003}), p.~\bibinfo{pages}{23}.

\bibitem[{\citenamefont{Aldana and Cluzel}(2003)}]{Aldana:03b}
\bibinfo{author}{\bibfnamefont{M.}~\bibnamefont{Aldana}} \bibnamefont{and}
  \bibinfo{author}{\bibfnamefont{P.}~\bibnamefont{Cluzel}},
  \bibinfo{journal}{Proc.\ Natl.\ Acad.\ Sci.\ USA}
  \textbf{\bibinfo{volume}{100}}, \bibinfo{pages}{8710} (\bibinfo{year}{2003}).

\bibitem[{\citenamefont{Greil and Drossel}(2005)}]{Greil:05}
\bibinfo{author}{\bibfnamefont{F.}~\bibnamefont{Greil}} \bibnamefont{and}
  \bibinfo{author}{\bibfnamefont{B.}~\bibnamefont{Drossel}},
  \bibinfo{journal}{Phys.\ Rev.\ Lett.} \textbf{\bibinfo{volume}{95}},
  \bibinfo{pages}{048701} (\bibinfo{year}{2005}).

\bibitem[{\citenamefont{Moreira and Amaral}(2005)}]{Moreira:05}
\bibinfo{author}{\bibfnamefont{A.~A.} \bibnamefont{Moreira}} \bibnamefont{and}
  \bibinfo{author}{\bibfnamefont{L.~A.~N.} \bibnamefont{Amaral}},
  \bibinfo{journal}{Phys.\ Rev.\ Lett.} \textbf{\bibinfo{volume}{94}},
  \bibinfo{pages}{218702} (\bibinfo{year}{2005}).

\bibitem[{\citenamefont{Kesseli et~al.}(2006)\citenamefont{Kesseli,
  R{\"a}m{\"o}, and Yli-Harja}}]{Kesseli:06}
\bibinfo{author}{\bibfnamefont{J.}~\bibnamefont{Kesseli}},
  \bibinfo{author}{\bibfnamefont{P.}~\bibnamefont{R{\"a}m{\"o}}},
  \bibnamefont{and}
  \bibinfo{author}{\bibfnamefont{O.}~\bibnamefont{Yli-Harja}},
  \bibinfo{journal}{Phys.\ Rev.\ E} \textbf{\bibinfo{volume}{74}},
  \bibinfo{pages}{046104} (\bibinfo{year}{2006}).

\bibitem[{\citenamefont{Paul et~al.}(2006)\citenamefont{Paul, Kaufman, and
  Drossel}}]{Paul:06}
\bibinfo{author}{\bibfnamefont{U.}~\bibnamefont{Paul}},
  \bibinfo{author}{\bibfnamefont{V.}~\bibnamefont{Kaufman}}, \bibnamefont{and}
  \bibinfo{author}{\bibfnamefont{B.}~\bibnamefont{Drossel}},
  \bibinfo{journal}{Phys.\ Rev.\ E} \textbf{\bibinfo{volume}{73}},
  \bibinfo{pages}{026118} (\bibinfo{year}{2006}).

\bibitem[{\citenamefont{Samuelsson and Socolar}(2006)}]{Samuelsson:06}
\bibinfo{author}{\bibfnamefont{B.}~\bibnamefont{Samuelsson}} \bibnamefont{and}
  \bibinfo{author}{\bibfnamefont{J.~E.~S.} \bibnamefont{Socolar}},
  \bibinfo{journal}{Phys. Rev. E} \textbf{\bibinfo{volume}{74}},
  \bibinfo{pages}{036113} (\bibinfo{year}{2006}).

\bibitem[{\citenamefont{Andrecut and Kauffman}(2006)}]{Andrecut:06}
\bibinfo{author}{\bibfnamefont{M.}~\bibnamefont{Andrecut}} \bibnamefont{and}
  \bibinfo{author}{\bibfnamefont{S.~A.} \bibnamefont{Kauffman}},
  \bibinfo{journal}{New J.\ Phys.} \textbf{\bibinfo{volume}{8}},
  \bibinfo{pages}{148} (\bibinfo{year}{2006}).

\bibitem[{\citenamefont{Klemm and Bornholdt}(2005{\natexlab{a}})}]{Klemm:05a}
\bibinfo{author}{\bibfnamefont{K.}~\bibnamefont{Klemm}} \bibnamefont{and}
  \bibinfo{author}{\bibfnamefont{S.}~\bibnamefont{Bornholdt}},
  \bibinfo{journal}{Phys.\ Rev.\ E} \textbf{\bibinfo{volume}{72}},
  \bibinfo{pages}{055101(R)} (\bibinfo{year}{2005}{\natexlab{a}}).

\bibitem[{\citenamefont{Flyvbjerg and Kj{\ae}r}(1988)}]{Flyvbjerg:88}
\bibinfo{author}{\bibfnamefont{H.}~\bibnamefont{Flyvbjerg}} \bibnamefont{and}
  \bibinfo{author}{\bibfnamefont{N.~J.} \bibnamefont{Kj{\ae}r}},
  \bibinfo{journal}{J.\ Phys.\ A: Math.\ Gen.} \textbf{\bibinfo{volume}{21}},
  \bibinfo{pages}{1695} (\bibinfo{year}{1988}).

\bibitem[{\citenamefont{Klemm and Bornholdt}(2005{\natexlab{b}})}]{Klemm:05b}
\bibinfo{author}{\bibfnamefont{K.}~\bibnamefont{Klemm}} \bibnamefont{and}
  \bibinfo{author}{\bibfnamefont{S.}~\bibnamefont{Bornholdt}},
  \bibinfo{journal}{Proc.\ Natl.\ Acad.\ Sci.\ USA}
  \textbf{\bibinfo{volume}{102}}, \bibinfo{pages}{18414}
  (\bibinfo{year}{2005}{\natexlab{b}}).

\bibitem[{\citenamefont{Elowitz and Leibler}(2000)}]{Elowitz:00}
\bibinfo{author}{\bibfnamefont{M.~B.} \bibnamefont{Elowitz}} \bibnamefont{and}
  \bibinfo{author}{\bibfnamefont{S.}~\bibnamefont{Leibler}},
  \bibinfo{journal}{Nature} \textbf{\bibinfo{volume}{403}},
  \bibinfo{pages}{335} (\bibinfo{year}{2000}).

\bibitem[{\citenamefont{Glass et~al.}(2005)\citenamefont{Glass, Perkins, Mason,
  Siegelmann, and Edwards}}]{Glass:05}
\bibinfo{author}{\bibfnamefont{L.}~\bibnamefont{Glass}},
  \bibinfo{author}{\bibfnamefont{T.~J.} \bibnamefont{Perkins}},
  \bibinfo{author}{\bibfnamefont{J.}~\bibnamefont{Mason}},
  \bibinfo{author}{\bibfnamefont{H.~T.} \bibnamefont{Siegelmann}},
  \bibnamefont{and} \bibinfo{author}{\bibfnamefont{R.}~\bibnamefont{Edwards}},
  \bibinfo{journal}{J.\ Stat.\ Phys.} \textbf{\bibinfo{volume}{121}},
  \bibinfo{pages}{969} (\bibinfo{year}{2005}).

\bibitem[{\citenamefont{Kaufman and Drossel}(2005)}]{Kaufman:05}
\bibinfo{author}{\bibfnamefont{V.}~\bibnamefont{Kaufman}} \bibnamefont{and}
  \bibinfo{author}{\bibfnamefont{B.}~\bibnamefont{Drossel}},
  \bibinfo{journal}{Eur.\ Phys.\ J.\ B} \textbf{\bibinfo{volume}{43}},
  \bibinfo{pages}{115} (\bibinfo{year}{2005}).

\bibitem[{\citenamefont{Strogatz}(2000)}]{Strogatz:00}
\bibinfo{author}{\bibfnamefont{S.~H.} \bibnamefont{Strogatz}},
  \emph{\bibinfo{title}{Nonlinear Dynamics and Chaos: With Applications to
  Physics, Biology, Chemistry, and Engineering}} (\bibinfo{publisher}{Westview
  Press}, \bibinfo{year}{2000}).

\bibitem[{\citenamefont{Kauffman}(1969)}]{Kauffman:69}
\bibinfo{author}{\bibfnamefont{S.~A.} \bibnamefont{Kauffman}},
  \bibinfo{journal}{J.\ Theor.\ Biol.} \textbf{\bibinfo{volume}{22}},
  \bibinfo{pages}{437} (\bibinfo{year}{1969}).

\bibitem[{\citenamefont{Derrida and Pomeau}(1986)}]{Derrida:86a}
\bibinfo{author}{\bibfnamefont{B.}~\bibnamefont{Derrida}} \bibnamefont{and}
  \bibinfo{author}{\bibfnamefont{Y.}~\bibnamefont{Pomeau}},
  \bibinfo{journal}{Europhys.\ Lett.} \textbf{\bibinfo{volume}{1}},
  \bibinfo{pages}{45} (\bibinfo{year}{1986}).

\bibitem[{\citenamefont{Pramila et~al.}(2006)\citenamefont{Pramila, Wu, Miles,
  Noble, and Breeden}}]{pramila:06}
\bibinfo{author}{\bibfnamefont{T.}~\bibnamefont{Pramila}},
  \bibinfo{author}{\bibfnamefont{W.}~\bibnamefont{Wu}},
  \bibinfo{author}{\bibfnamefont{S.}~\bibnamefont{Miles}},
  \bibinfo{author}{\bibfnamefont{W.~S.} \bibnamefont{Noble}}, \bibnamefont{and}
  \bibinfo{author}{\bibfnamefont{L.~L.} \bibnamefont{Breeden}},
  \bibinfo{journal}{Genes \& Dev.} \textbf{\bibinfo{volume}{20}},
  \bibinfo{pages}{2266} (\bibinfo{year}{2006}).

\end{thebibliography}

\end{document}